\documentstyle[12pt,psfig]{article}
\begin{document}

\begin{titlepage}
\begin{flushright}
CUPP-97/6 \\
SINP/TNP/97-17\\
astro-ph/9807100
\end{flushright}
\begin{center}
{\large {\bf Oscillation Effects On Neutrinos From The Early Phase Of a
Nearby Supernova}}
\end{center}
\begin{center}
  {\large Debasish Majumdar$^1$, 
        Kamales Kar$^2$, 
        Alak Ray$^{3,4}$, \\ 
        Amitava Raychaudhuri$^{1,6}$, 
        Firoza K. Sutaria$^{3,5}$}
\end{center}

\begin{flushleft}
$^1$Department of Physics, University of Calcutta,\\ 
92 Acharya Prafulla Chandra Road, Calcutta 700 009, India \\
$^2$Saha Institute of Nuclear Physics, 1/AF Bidhannagar, Calcutta
700 064, India \\
$^3$Tata Institute of Fundamental Research, Homi Bhabha Road,\\
Mumbai (Bombay) 400 005, India \\
$^4$Code 661, LHEA, NASA/Goddard Space Flight Center, Greenbelt,
MD 20770, U.S.A. \\
$^5$Inter University Centre for Astronomy \& Astrophysics, Pune 411
007 India \\
$^6$Department of Physics, Oklahoma State University, Stillwater, OK
74078, U.S.A.
\end{flushleft}

\begin{abstract}

Neutrinos emitted during stellar core collapse leading to a supernova
are primarily of the electron neutrino type at source which may undergo
oscillation between flavor eigenstates during propagation to an
earth-bound detector. Although the number of neutrinos emitted during
the pre-bounce collapse phase is much smaller than that emitted in the
post-bounce phase (in which all flavors of neutrinos are emitted), a
nearby supernova event may nevertheless register a substantial number
of detections from the pre-bounce phase at SuperKamiokande (SK) and the
Sudbury Neutrino Observatory (SNO).  The calorimetric measurement of
the supernova neutrino fluence from this stage via the charge current
and neutral current detection channels in SNO and the corresponding
distortion of detected spectrum in SK over the no-oscillation spectrum,
can probe information about neutrino mass difference and mixing which
are illustrated here in terms of two- and three-flavor oscillation
models.

\end{abstract}
\end{titlepage}

\section*{Introduction}

Large underground experiments with the main purpose of detecting solar
neutrinos are already on-line ({\it e.g.} SuperKamiokande, hereafter SK
- see Totsuka 1990) or shortly expected to be operational (Sudbury
Neutrino Observatory, hereafter SNO 1987). The latter experiment,
because of its capability to detect neutrinos {\it via} both
charged-current as well as neutral-current detection channels is
expected to probe the role of neutrino oscillation in the reduction of
solar neutrino flux from its ``standard" value (Bahcall 1989, SNO
1987). These experiments will also be able to easily detect neutrinos
from a supernova explosion anywhere in the Milky Way galaxy. If a core
collapse supernova occurs sufficiently nearby (typically 1-2 kpc away),
then they should detect a substantial number of neutrinos even from the
stellar core collapse phase (Sutaria \& Ray 1997 (SR)) which are far
less copious than the neutrinos from the post-bounce phase (Burrows
1990; Burrows et al. 1990, Totani et al.  1997).  Neutrinos from the
core-collapse phase, like the solar neutrinos, are mainly of the
electron neutrino type - the product of continuum electron capture on
free protons and neutron-rich nuclei present in the stellar core. Thus,
if the calorimetric measurement of the neutrino fluence in the neutral
current channel in SNO -- which is neutrino flavor insensitive --
exceeds the energy integrated neutrino fluence in the charged current
channel, after accounting for corresponding detection efficiencies,
then this would be a clear signal of neutrino flavor oscillation during
propagation. Such an event and its detection will probe the neutrino
mass and mixing parameters in a very different regime than that
possible from measurement of solar, atmospheric and accelerator
neutrinos.

Totani (1998) has considered the possibility of neutrino mass
determination by using arrival time sequence analysis at different
energies of the abundant post-bounce (anti)neutrinos in the
Super-Kamioka detector.  
This proposed method of neutrino mass determination
has been argued to remain valid
in the case of almost degenerate hierarchy of
neutrino masses ($m_{\nu_e} \simeq m_{\nu_{\mu}} \simeq
m_{\nu_{\tau}}$) even with neutrino flavour oscillations.
In this paper we consider the complementary
information that can be probed from supernova about neutrino mass {\it
differences} and other associated 
parameters due to neutrino oscillation between 
different flavor eigenstates in a nearly degenerate mass hierarchy by
using a combination of spectra and total number of events in certain
channels provided by these two different experiments.  Totani (1998)
indicates that $m_{\nu_e}$ of $\sim$ 3 eV can be probed by a future
galactic supernova, whereas we show here that a mass {\it difference}
scale $\sqrt \Delta$ will be probed at 10$^{-9}$ eV should the core
collapse phase neutrinos be detected by SK and SNO. 

\section*{Oscillation of Neutrinos From Stellar Core Collapse Phase}

The recent SuperKamiokande announcement of the discovery of neutrino
mass and oscillation in their atmospheric $\nu$ data (Kajita 1998) is a
milestone in the physics of neutrinos with wide implications for
astrophysics and particle physics.  It has, of course, been  widely
held for some time that oscillations of $\nu_e$ to a neutrino of a
different flavor may hold the key to a solution of the long-standing
solar neutrino problem. Encouraged by these developments, we first
discuss the effect on the SN signal of the oscillation of $\nu_e$ with
one other neutrino type in some detail. This study indicates that for
certain ranges of the mass-difference and the mixing angles one gets
considerable depletion in the number of detected neutrinos.  The study
is carried out for two different sets of zero age main sequence stellar
masses corresponding to different sets of initial conditions prior to
core collapse, {\it i.e.} for a 15M$_\odot$ and a 25M$_\odot$ star,
though we present results mostly for the former case.  The LEP
experiments at CERN have established that there are three light
neutrino flavors. In view of this, we also extend the analysis to the
case of three-neutrino mixing. In this scenario, we choose the three
neutrino masses so that one of the oscillation lengths is of the order
of the sun-earth distance while the other is of a distance
corresponding to a galactic supernova in the solar neigbourhood.

The core of a massive star collapses under its own gravity once the
nuclear burning stops and the pressure support from degenerate
electrons is reduced due to electron capture. Neutrinos are emitted
from the supernova at two stages: the first during collapse in a burst
of about ten milliseconds from neutronization through electron capture
until neutrino trapping sets in and the second one in the hotter
post-bounce era in the form of thermal neutrinos of all three flavors
in a time-scale of a few seconds. In this report we shall concern
ourselves with the first stage though the flux of post-bounce neutrinos
is higher.

Expected number of detections in SNO and SK have been calculated for
neutrinos from a typical 15M$_\odot$ star undergoing core collapse at a
distance of 1 kpc during its neutronization phase (SR). The neutrino
spectra observable by these experiments, for a realistic range of
nuclear physics input and stellar masses on the main sequence were also
given.  Neutrinos which are emitted before they undergo inelastic
scattering and trapping by overlying stellar matter carry with them
information about the physical conditions within the core as well as
its nuclear configuration and hence their detection can be important to
understand the supernova dynamics.  SR use a one-zone collapse code
(Ray et al. 1984) to generate the energy spectrum of the emitted
neutrinos and consider the electron capture on both free protons and
heavy nuclei (in the {\it fp} shell) the abundance distribution of
which are self consistently determined with the evolution of
thermodynamic conditions as collapse proceeds using an analytical
equation of state of warm dense matter due to Bethe et al. (1979) as
modified by Fuller (1982).

SR (see also Sutaria 1997) presented a cumulative neutrino fluence till
the stellar core reaches a mean density of $2.4\times 10^{11}$
gm/cm$^3$ beyond which stage the neutrinos begin to be trapped and
undergo inelastic scattering. For completeness, we present in Fig. 1
the results for both the 15M$_\odot$ (solid line)
and 25M$_\odot$ (dashed line) cases taking the distance to be 1 kpc.
The spectra depicted are a sum of contributions from electron capture
on both free protons and heavy nuclei, while SR presented the separate
contributions of these components for the 15 M$_{\odot}$ case.  The
fluence is folded with the cross-section for the charge current (CC)
reaction in the D$_2$O detector at SNO -- $\sigma_{c.c.} = 1.7 \times
10^{-44}{\rm cm}^2 (E_\nu - 1.44)^{2.3}$ (Burrows, 1990) -- to predict
the number of neutrinos detectable as a function of the energy. These
results are shown for the 15M$_\odot$ star in Fig. 2
(upper panel - solid curve).  In addition,
the neutrinos also undergo neutral current interactions but for this
process the detector cannot give incident $\nu $-energy information and
only calorimetric measurements are possible.  We have also folded the
fluence with the neutral current (NC) cross-section -- $\sigma_{n.c.} =
0.85 \times 10^{-44}{\rm cm}^2 (E_\nu - 2.2)^{2.3}$ (Burrows 1990) --
to obtain a result which is not affected by oscillations since the NC
interaction is flavor-blind. We find that the total ({\it i.e.,} energy
integrated) signal due to the neutral current reaction is less than
that for the charged current case due to differing detection
efficiencies.  It is seen that the ratio of the total NC signal and the
corresponding CC signal at SNO -- which is relatively insensitive to
the details of the initial star -- is a useful measure for oscillations
(see later).

The SuperKamiokande detector uses 32 ktons of light water in which
electrons scattered by $\nu_e$ -- through both charged and neutral
current interactions -- are detected {\it via} \u{C}erenkov radiation.
In this case, the cumulative fluence is folded with the $\nu_e - e^-$
scattering cross-section -- $ \sigma = 0.94 \times 10^{-43} {\rm cm}^2
(E_\nu /10 {\rm MeV})$ (Sehgal, 1974). The results for this detector
are shown for the 15M$_\odot$ star in Fig. 2
(lower panel - solid curve).

In this paper we estimate the effect on both the SNO and SK signals
when neutrino oscillations are operative. We consider oscillations of
the electron neutrinos to other sequential neutrinos: {\it i.e.,}
$\nu_{\mu}$ or $\nu_{\tau}$. (We make some comments about oscillation
to sterile -- {\it non-interacting} -- neutrinos towards the end). In
the two-flavor case, the probability of an electron neutrino of energy
$E_{\nu}$ to oscillate to a neutrino of a different type -- $\nu_x, x
\equiv \mu$ or $\tau$ -- after the traversal of a distance $L$ is:

\begin{equation}
P_{\nu_e \rightarrow \nu_x} = \sin^2(2\theta) \sin^2 \left (\frac {\pi
L} {\lambda} \right )
\label{ex}
\end{equation}
where $\theta$ is the mixing angle and the oscillation length is given
in terms of the mass-squared difference $\Delta$ by:  
\begin{equation}
\lambda = 2.47 \left (\frac
{E_\nu} {\rm MeV} \right ) \left (\frac {\rm eV^2} {\Delta}\right
)\; {\rm meter}
\end{equation}
From probability conservation: $P_{\nu_e \rightarrow \nu_e} =  1 -
P_{\nu_e \rightarrow \nu_x}$.

Consider now the effect of oscillations on the neutrino signal from the
early phase of a galactic supernova. At SNO, the $\nu_{\mu}$ or
$\nu_{\tau}$ generated by oscillations cannot interact {\it via} the
charged current. The prediction for this signal is obtained by folding
$P_{\nu_e \rightarrow \nu_e}$ with the fluence and the charged current
neutrino absorption cross-section. It undergoes a depletion shown for
some typical cases -- $\Delta = 1 \times 10^{-18}$ eV$^2$ and $\theta $
= 30$^o$ (long-dashed curve) and $\Delta = 2.5 \times 10^{-18}$ eV$^2$
and $\theta $ = 35$^o$ (small-dashed curve) -- in Fig. 2.
The oscillatory behavior of neutrinos,
though modulated by the fluence, is clearly perceptible and change with
$\Delta$ as expected. In particular, for the chosen $\Delta$ the signal
at the high energy end suffers a strong depletion.

   The smallness of our choice of $\Delta$ might call for a remark.
The energy and length scales associated with supernova neutrinos
provide a unique window for very small mass splittings -- a point noted
earlier by Reinartz and Stodolsky (1985). Small though this may appear
at first sight, it is only five orders of magnitude smaller 
than that for $K^o$s.

The $\mu$ or $\tau$ neutrinos contribute to the SuperKamioka signal
only through neutral current interactions for which the cross-section
is $\sigma(\nu_x \rightarrow e) = 1.6 \times 10^{-44} {\rm cm}^2 (E_\nu
/{\rm 10 MeV})$, $x =\mu$ or $\tau$ (Kolb et al. 1987). The effect on
the detected signal in this case is presented in Fig. 2
(lower panel). The oscillatory behavior of
the signal in this case is again apparent, more so since the SK signal
is much broader than that of SNO.

In SNO, the $\nu_\mu$ and $\nu_\tau$ neutrinos will not induce charged
current reactions but will undergo neutral current interactions with
full strength.  Since the neutral current signal at SNO is immune to
oscillations, the ratio, $R_{SNO}$, of the calorimetric detection of
the neutrino fluence via the NC channel to the total (energy integrated
fluence) detection via the CC channel (5 MeV $\leq E_{\nu} \leq $ 25
MeV) is a useful probe for oscillations. Some results for this ratio
are presented in Table 1.  For comparison we have shown the results for
$R_{SNO}$ without and with oscillation for both $15M_{\odot}$ and
$25M_{\odot}$ stars. It is seen that the value of $R_{SNO}$ is not
sensitive to the typical examples of stellar collapse shown here, which
includes a combination of initial conditions as reflected in the zero
age main sequence mass of the pre-supernova star, matrix elements of
the electron capture on heavy nuclei etc. -- it remains fixed at 0.434
while going from $15M_{\odot}$ to $25M_{\odot}$.  Therefore, if
$R_{SNO}$ is observed to be significantly larger than the predicted
no-oscillation value then this difference cannot be attributed to the
range of variations expected from astrophysical and nuclear physics
grounds. The relative strengths of the Supernova neutrino signal from
electron capture on free protons and on heavy nuclei is determined by
the dynamics and thermodynamics of the initial star. It is seen from
Table 1 that the range of variation of $R_{SNO}$ due to capture only on
protons or only on heavy nuclei (which encompass any intermediate
possibility) is not large -- {\it e.g.} for the $15M_{\odot}$ case, it
varies from 0.444 to 0.396 -- and, in particular, cannot mimic the
effect of oscillation which can yield values as large as 0.776.
Therefore a measured value of $R_{SNO}$ different from the theoretical
prediction will be indicative of neutrino oscillations. In Fig. 3
we present contours of constant $R_{SNO}$ in the $\Delta -
\theta$ plane. The shape of these contours can be readily explained
using eq.(~\ref{ex}).

\section*{The Three Flavor Case}

Now let us turn to the case of three flavors. Though we restrict
ourselves to three flavors for simplicity, the experimental indications
point towards four or more neutrino types. This is because the
explanation of the solar, atmospheric, and LSND $\nu $ results within
the neutrino oscillation framework requires vastly different mass
differences. In view of the findings of the LEP experiments, this
would require the introduction of one or more sterile neutrinos.

The two flavor case can always be recovered from the three flavor one
by a suitable choice of the mixing angles.  But the three flavor
scenario has some additional features which merit examination.  The
general expression for the probability for a $\nu_\alpha$ to oscillate
to $\nu_\beta$ after traversing a distance $L$ is given by (Goswami et
al. 1997)
\begin{equation}
P_{\nu_\alpha\nu_\beta} = \delta_{\alpha\beta} -4\sum_{j>i} 
U_{\alpha i}U_{\beta i}U_{\alpha j}U_{\beta j} \sin^2 \left (\frac 
{\pi L} {\lambda_{ij}} \right )
\label{pab}
\end{equation}
Here the flavor basis is identified by greek indices  while the mass
eigenstates are represented by roman indices.  $L$ is the distance from
the detector to the source and $\lambda_{ij} = 2.47 \left( {E_\nu}/
{\rm MeV}  \right ) \left ({\rm eV^2} / {\Delta_{ij}}\right
)$ meter, is a characteristic oscillation length. In the above,
$\Delta_{ij} = m_j^2 - m_i^2$.  The matrix $U$, which relates the
flavor eigenstates to the mass eigenstates, can be represented in the
three-flavor case in terms of three mixing angles $\theta_{ij} $ as:
\begin{equation}
U = \left ( \begin{array}{ccc} 
c_{12}c_{13} & s_{12}c_{13}c_{23} - s_{13} s_{23} & 
c_{13}s_{12}s_{23} +  s_{13}c_{23} \\
-s_{12} & c_{12}c_{23} & c_{12}s_{23} \\
-s_{13}c_{12} & -s_{13}s_{12}c_{23} - c_{13} s_{23} & 
-s_{12}s_{13}s_{23} +  c_{1s}c_{23} \end{array} \right )
\label{mat}
\end{equation}
where $c_{ij} = \cos\theta_{ij}$ and $s_{ij} = \sin\theta_{ij}$. Since
we are not interested in CP-violation in the neutrino sector, $U$
has been chosen real. The flavor states are: $\alpha = $ 1
($e$), 2 ($\mu$), and 3 ($\tau$).

In line with the discussion earlier, we consider one of the mass
splittings $\Delta_{23}$ to be around $10^{-18}$eV$^2$ so that the the
oscillation length $\lambda_{23}$ for neutrinos of energies in the 10
MeV range is around 1 kpc. We choose the other splittings $\Delta_{12}
\simeq \Delta_{13} = 10^{-10}$ eV$^2$ so that $\lambda_{12} \simeq
\lambda_{13}$ corresponds roughly to the earth-sun distance (assuming
that the discrepancy of measured and astrophysically predicted solar
neutrino fluxes is due to neutrino flavor oscillations on such length
scales).
In the present case of a galactic Supernova, we set $L$ to 1 kpc ($\gg
\lambda_{12} \simeq \lambda_{13}$) and one has:
\begin{eqnarray}
P_{\nu_e \rightarrow \nu_e} &=& 1 - 2c_{13}^2c_{12}^2 + 
2c_{13}^4c_{12}^4  \nonumber \\
&-& 4(c_{13}s_{12}c_{23} - s_{13}s_{23})^2 (c_{13}s_{12}s_{23} + 
s_{13}c_{23})^2 \sin^2 \left (\frac {\pi L} {\lambda_{23}} \right )
\label{ee}
\end{eqnarray}
We do not present the explicit forms of 
$P_{\nu_e \rightarrow \nu_\mu}$ and $P_{\nu_e \rightarrow \nu_\tau}$
which can be readily obtained.

As noted earlier, the $\nu_{\mu}$ and $\nu_{\tau}$ cannot produce the
charged current signal at SNO.  Due to three-flavor oscillation, this
signal undergoes a depletion shown for the purposes of illustration for
a typical choice of mixing angles in Fig. 4 (upper
panel - dashed line) where we have chosen $\theta_{12} = 6.5^o$,
$\theta_{13} = 25^o$ and $\theta_{23} = 10^o$ and find a reduction in
the signal of $\sim 30 \%$. The results for SuperKamioka are similar
and are presented for the same choice of neutrino mixing parameters in
Fig. 4 (lower panel).

In Fig. 4, the results for the parameter set
$\theta_{12} = 6.5^o$, $\theta_{13} = 25^o$ and $\theta_{23} = 10^o$
have been presented.  We have examined the effect of varying all the
three angles and have found that the reduction for both SNO and SK is
only mildly sensitive to $\theta_{23}$ and depends more on
$\theta_{12}$ and $\theta_{13}$. For example, if $\theta_{12}$ is
increased from 6.5$^o$ to 15$^o$ while the other mixing angles are
unchanged, then the percentage reduction for SNO (SK) increases from
$\sim 30\%$ ($\sim 26\%$) to $\sim 37\%$ ($\sim 31\%$).  Similarly, a
change of $\theta_{13}$ from 25$^o$ to 10$^o$, keeping the other two
angles fixed, results in a drop of the reduction to $\sim 17
\%$ ($\sim 16\%$). These results can be readily understood from eq.
(~\ref {ee}) from where it can be concluded that so long as the
$\theta_{ij} $ mixing angles are small, the leading $\theta_{12},
\theta_{13}$ contributions depend quadratically on the angles while the
leading $\theta_{23}$ dependence is multiplied by a quartic product of
the other angles.

If the neutrino mass-squared difference $\Delta_{23}$ is much smaller
than the chosen $10^{-18}$ eV$^2$, then the corresponding oscillation
length will be larger than $L$ and the effect of this mode will not be
seen -- {\it e.g.,} the last term in eq. (~\ref {ee}) will be absent. On
the other hand, if $\Delta_{23}$ is much bigger than the chosen value
then the oscillatory term in these equations -- $ \sin^2 \left ({\pi
L}/ {\lambda_{23}} \right )$ -- will be averaged to $\frac{1}{2}$. Thus
the detection of a distorted spectrum of neutrinos from the core
collapse phase of a supernova by SuperKamioka and SNO would probe a
different range of the mass-mixing parameters (than the solar neutrino
parameters) - which is determined by the distance to the supernova - an
observable quantity which can be astronomically determined {\it a
posteriori}. If the distance to the supernova is however much larger
than 1 kpc, the number of detectable neutrinos from this phase may be
too small to effectively probe the corresponding mass-mixing region of
the neutrino flavors.  An undistorted spectrum on the other hand, would
be a null experiment since the parameters as discussed above may be
different from what is being probed via such experiments.

The neutrinos produced in the pre-bounce stage are all of the
electron-type and are produced in a time span of the order of tens of
milliseconds. Through oscillations, neutrinos of other flavors are
generated. On the other hand, in the post-bounce era, neutrinos and
anti-neutrinos of all three flavors are produced. If the difference in the
time of travel, $\Delta T$, due to neutrino mass splitting is so large as
to wipe out the time gap between the signals from the pre- and post-bounce
phase then the detection of $\nu_{\mu}$ or $\nu_{\tau}$ in the observed
beam cannot be unequivocally attributed to oscillations.  However, for the
mass ranges that would be probed by a SN at 1 kpc, $\Delta T$ is much
smaller than the duration of the stellar core collapse ($\simeq$ 10 ms). 
This is readily seen as follows. In order to avoid irrelevant
complications consider the two-flavor scenario. Here we have:  
$$ 
\Delta T
= \displaystyle\frac {L} {c} \left[ \frac {1} {\beta_1} - \frac {1}
{\beta_2} \right ] 
$$ 
Here, $E/m = 1/\sqrt {1 - \beta^2}$ whence $(\beta_2^2 - \beta_1^2) =
m_1^2/E_1^2 - m_2^2/E_2^2 \simeq \Delta/E^2$.  Choosing $\Delta =
10^{-18}$ eV$^2$, taking $E = 10$ MeV ($\beta \sim 1$), and with $L =$
1 kpc, $\Delta T \simeq 10^{-21}$ sec.

\section*{Discussion and Conclusions}

In this work, we have considered the oscillation of $\nu_e$ to other
sequential neutrinos. If instead, the oscillation is to sterile
neutrinos -- with no coupling to the electroweak gauge bosons -- then
the following changes will occur. In SNO, both the charged {\it and}
neutral current interactions will suffer depletion and thus the ratio
$R_{SNO}$ (see Table 1) will be almost unchanged. However, in SK the
sterile neutrinos will have no interaction whatsoever (unlike the NC
interactions of the sequential neutrinos) and hence the signal will be
reduced even further. We hope to make a detailed, comparative
assessment of the two scenarios in subsequent work.

The mass square differences we have considered -- $10^{-18}$ eV$^2$ --
dictated by the energy and length scales that characterize the
situation are indeed very small. For a comparison note that the recent
announcement of the discovery of neutrino oscillation in the
atmospheric $\nu $ data by SuperKamiokande relies on neutrinos produced
in the atmosphere with energies in the GeV scale and with path lengths
ranging from tens of kilometres (downward going $\nu$s) to around ten
thousand kilometres (upward going $\nu$s) (Kajita 1998). They are
therefore sensitive to mass differences around $10^{-2}$ eV$^2$ which
may be tested by the long baseline accelerator experiments. Solar
neutrino experiments with typical MeV energies (Bahcall 1989) may
signal mass differences in the $10^{-5}$ eV$^2$ (MSW) to the $10^{-10}$
eV$^2$ (vacuum oscillation) range. The results of the LSND experiment
(Athanassopoulos et al., 1995) are indicative of $\nu_{\mu}
\leftrightarrow \nu_e$ oscillation with mass differences of the order
of eV$^2$. In view of these results and the current availability of
large neutrino detectors, supernova neutrinos provide a means to probe
a new scale of mass differences.

Although the {\it a priori} event rate of such nearby supernova
explosion may be argued to be low, (Strom (1994) has given the
historical supernova rate in the galaxy to be 5.7 per century while
Cappellaro {\it et al.} (1997), based on SN rates in external galaxies
give this rate for a galaxy similar to ours as 2.2 $\pm$ 1.3
SN/century), it is difficult to predict with confidence the time of the
next such occurrence for such a low event rate phenomenon. In the Orion
and other constellations there are a number of super-giant stars within
about 500 pc of the sun which have spectral types and absolute
magnitudes similar to the progenitor of Supernova 1987A -- Sanduleak
69$^o$-202 -- which apart from the classical candidate Betelgeuse -- a
red super-giant 200 pc away -- are potential core-collapse supernova
progenitors.  It is well-known that SN of type II and Ib/Ic occur in or
near the spiral arms of the galaxy from massive star progenitors.
Therefore if the progenitors are uniformly strung across 
spiral arms, the relative number of supernovae within a distance r
of the sun (normalised to the total galactic rate) would scale as the
ratio of the cumulative lengths of the spiral arms within a distance r
from the sun compared to the total cumulative lengths of all spiral
arms of the galaxy.  Significant parts of three spiral arms of the
galaxy (which are predominantly the sites of type II and Ib/Ic
explosions) namely, Cygnus, Saggitarius and the Perseus arms, lie
within about 1-2.5 kpc of the sun.  The cumulative lengths of all
spiral arms in the galaxy can be compared with the 
total lengths of the spiral arms within
$\sim$ 2 kpc of the sun from the Cygnus,
Saggitarius and Perseus arms (see e.g. the figures from
Taylor and Cordes (1993) and 
Yadigaroglu and Romani (1997)) - the latter being roughly
10\% of the former.  Thus the rate of nearby SNe (within
about 2 kpc of the sun - so that both Super-Kamioka and SNO would be
effective detectors of collapse phase neutrinos) would be about 10\%
of the galactic SN rate.  Many of the historical SNe in the last
millennium in our galaxy have taken place within 2-3 kpc of the sun.
Therefore it is reasonable to expect from the above scaling arguments
that the time interval between nearby supernovae (roughly at 1-2 kpc
distance - at which SuperKamioka and SNO will still be able to detect
the infall neutrinos) may be once every two to several hundred years.
Kepler's SN occurred in 1604, and there has been short intervals of time
between supernovae such as the type II Crab SN and SN 1181.  The
evolutionary status of the core of a massive star may remain uncertain
within several thousands of years from the observed surface properties
of a super-giant.  As in the case of the progenitor of SN1987A, some of
the super-giant stars in the solar neigbourhood may in reality be close
to the stage of explosion and might eventually provide an opportunity
during the projected lifetime of SK ($\simeq$ 50 yrs) to experimentally
constrain aspects of stellar core collapse and neutrino properties
apart from the characteristics of explosion models.

In summary, the neutrino signal from the early phase of a nearby
Supernova can undergo significant and detectable modification if
neutrino oscillations are operative. Such a distortion of the signal
may provide a means to shed new light on neutrino masses and mixing. In
particular, it is a unique probe of mass differences ($\Delta \sim
10^{-18}$ eV$^2$) much smaller than those relevant to the solar
neutrino problem (typically $\sim 10^{-10}$ eV$^2$ for vacuum
oscillations) or the atmospheric neutrino anomaly ($\sim 10^{-2}$
eV$^2$).

\section*{Acknowledgements}

The work of D.M., K.K., and A.Raychaudhuri is partially supported by
the Eastern Center for Research in Astrophysics, India. A.Raychaudhuri
also acknowledges a research grant from the Council of Scientific and
Industrial Research, India. A. Ray was a Senior Research Associate of
the U.S. National Research Council at NASA/GSFC during part of this 
work. He also thanks the
Director and the Coordinators of the Supernova program at the 
Institute of Theoretical Physics, Santa Barbara for their hospitality. 
The authors wish to thank Dr.  Sukhendusekhar
Sarkar for a valuable suggestion.

\newpage
Table 1: The ratio, $R_{SNO}$, of the  neutral current detections
to the charge current detections at SNO (5 MeV $\leq E_{\nu_e} \leq$ 25
MeV) with and without two flavor neutrino oscillations
($\theta = 30^o$ and $\Delta = 1 \times 10^{-18}$).
\vskip 0.5 true cm
\begin{center}
\begin{tabular}{|c|ccc|ccc|}
\hline
\hline
Star Mass  &
\multicolumn{3}{|c|} {R$_{SNO}$ without oscillation}&
\multicolumn{3}{|c|} {R$_{SNO}$ with oscillation}  \\ \cline{2-7}
& free protons & heavy nuclei & total & free protons &
heavy nuclei & total \\
\hline
15 M$_{\odot}$ &  0.444   & 0.396   & 0.434  & 0.840    & 0.577
&0.776  \\ \hline
25 M$_{\odot}$ &  0.440   & 0.394   & 0.434  & 0.787    & 0.581
&0.754  \\
\hline
\end{tabular}
\end{center}

\newpage
\pagestyle{empty}
\begin{figure}
\par
\centerline{
\psfig{figure=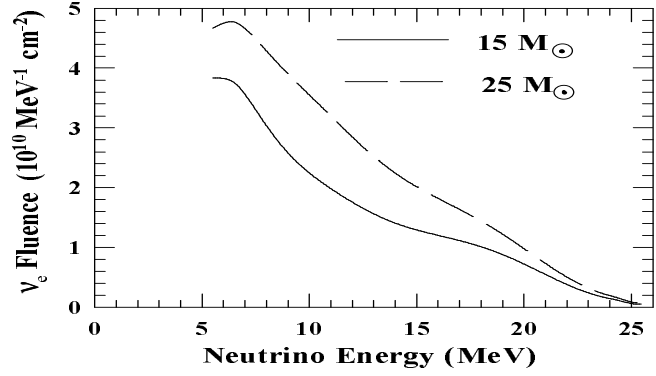,height=8in,width=8in,angle=0}
}
\par
\caption {Cumulative incident neutrino fluence till
mean stellar density $\rho$ reaches $2.42 \times 10^{11} \rm gm cm^{-3}$
for 15M$_\odot$
(solid line) and 25M$_\odot$ (dashed line) stars 1 kpc away assuming no
flavor oscillations.}
\end{figure}
\newpage
\pagestyle{empty}
\begin{figure}
\par
\centerline{
\psfig{figure=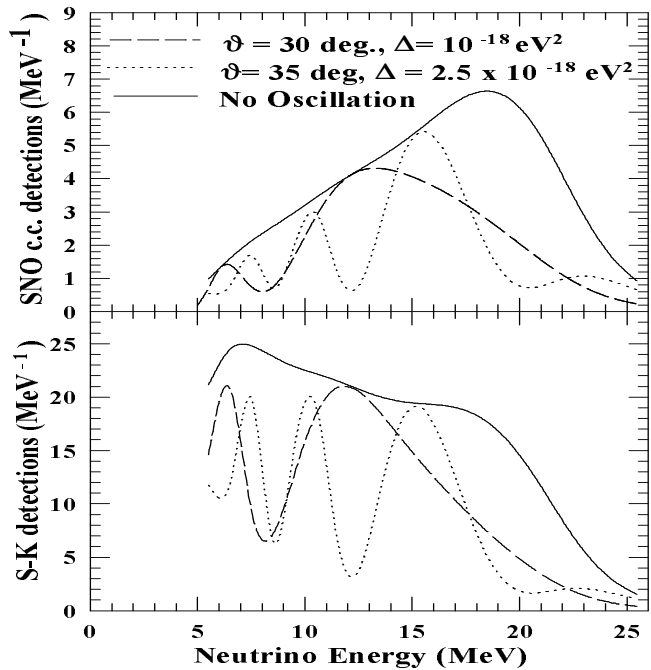,height=8in,width=8in,angle=0}
}
\par
\caption {Charged current neutrino signal expected at SNO (upper panel) and
SuperKamioka (lower panel) with and without oscillations for a
15M$_\odot$ star 1 kpc away.  The solid line represents the
no-oscillation case while the large-dashed (small-dashed) line
corresponds to two-flavor oscillations with $\Delta = 1 \times
10^{-18}$ eV$^2$ and $\theta $ = 30$^o$ ($\Delta = 2.5 \times 10^{-18}$
eV$^2$ and $\theta $ = 35$^o$) which corresponds to $R_{SNO}$ (see
later) = 0.776 (0.895).  }
\end{figure}
\pagestyle{empty}
\begin{figure}
\par
\centerline{
\psfig{figure=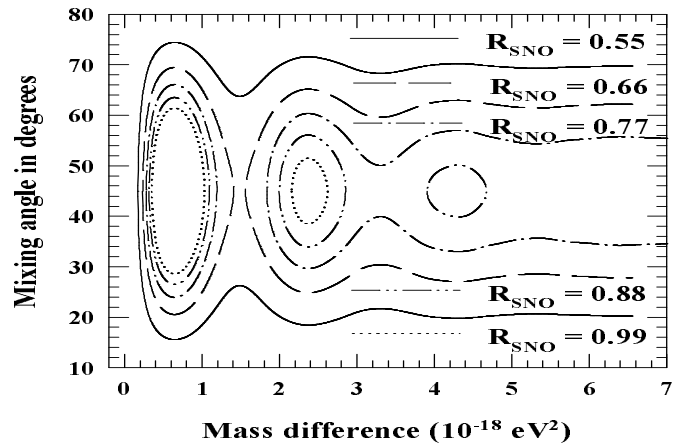,height=8in,width=8in,angle=0}
}
\par
\caption {Contours of constant $R_{SNO}$ -- the ratio of the NC signal to the
energy integrated CC signal at SNO -- in the $\Delta - \theta$ plane.
No neutrino oscillation corresponds to $R_{SNO}$ = 0.434.  }
\end{figure}
\pagestyle{empty}
\begin{figure}
\par
\centerline{
\psfig{figure=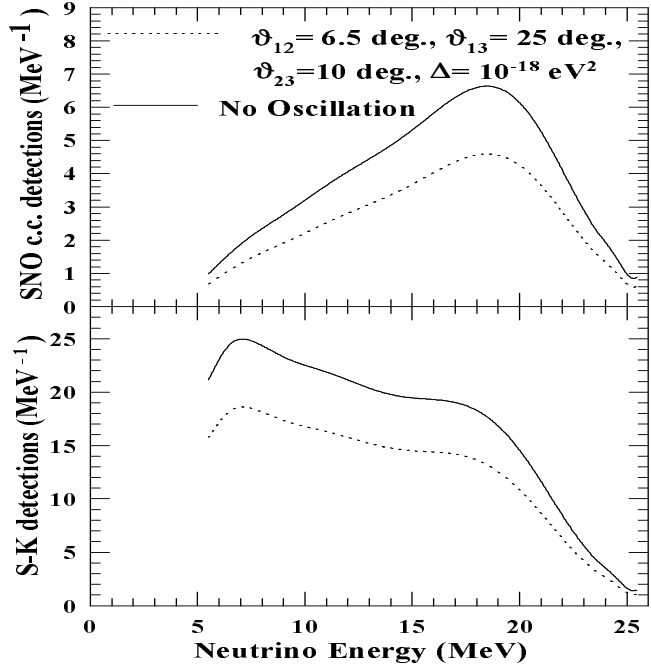,height=8in,width=8in,angle=0}
}
\par
\caption {Charged current neutrino signal expected at SNO
(upper panel) and at SuperKamioka (lower panel)
with and without oscillations in the three-flavor case for a 15M$_\odot$
star 1 kpc away.  The solid line represents the
no-oscillation case while the dashed line corresponds to three-flavor
oscillations with $\Delta_{23} = 1 \times 10^{-18}$ eV$^2$ and
$\theta_{12} $ = 6.5$^o$, $\theta_{13} $ = 25$^o$, and $\theta_{23} $ =
10$^o$. This choice corresponds to $R_{SNO}$ =  0.629.} 
\end{figure}

\end{document}